\documentclass[12pt]{article}
\usepackage[a4paper, margin=1in]{geometry}
\usepackage{graphicx} 
\usepackage{mathptmx}
\usepackage[T1]{fontenc}
\usepackage{authblk}
\usepackage{titlesec}
\usepackage{etoolbox}
\usepackage{setspace}
\usepackage{abstract}
\usepackage{amsmath}
\usepackage{array}
\usepackage{booktabs}
\usepackage{moreverb,url}
\usepackage{lineno}
\usepackage{hyperref}
\usepackage{xcolor}
\usepackage[superscript,biblabel]{cite}
\usepackage{doi}
\usepackage{tikz}

\titleformat{\section}
{\normalfont\bfseries\uppercase}
{\thesection.}{0.5em}{} 

\titleformat{\subsection}
{\normalfont\bfseries}
{\thesubsection.}{0.5em}{} 

\titleformat{\subsubsection}
{\normalfont\bfseries\itshape}
{\thesubsubsection.}{0.5em}{}

\makeatletter
\patchcmd{\@maketitle}{\begin{center}}{\begin{flushleft}}{}{}
\patchcmd{\@maketitle}{\begin{tabular}[t]{c}}{\begin{tabular}[t]{@{}l}}{}{}
\patchcmd{\@maketitle}{\end{center}}{\end{flushleft}}{}{}
\patchcmd{\@maketitle}{\LARGE}{\large}{}{}
\makeatother

\title{\textbf{Estimation of the Acoustic Field in a Uniform Duct with Mean Flow using Neural Networks}}
\date{}

\author{\raggedright
\textbf{D. Veerababu} \newline
\textit{Department of Electrical Engineering} \newline
\textit{Indian Institute of Science, Bengaluru, India - 560012} \newline
\textit{veerababudha@iisc.ac.in} \and

\raggedright\textbf{Namra Quasim} \newline
\textit{Department of Electronics and Communication Engineering} \newline
\textit{R.V. College of Engineering, Bengaluru, India - 560059} \newline
\textit{namraquasim.ec19@rvce.edu.in} \and

\raggedright\textbf{Prasanta K. Ghosh} \newline
\textit{Department of Electrical Engineering} \newline
\textit{Indian Institute of Science, Bengaluru, India - 560012} \newline
\textit{prasantg@iisc.ac.in}
}

\begin{document}
\maketitle 
\pagebreak

\newcommand{\blueline}{\raisebox{2pt}{\tikz{\draw[-,blue,solid,line width = 1.5pt](0,0) -- (10mm,0);}}}
\newcommand{\bluedotline}{\raisebox{2pt}{\tikz{\draw[-,blue,dotted,line width = 1.5pt](0,0) -- (10mm,0);}}}
\newcommand{\redline}{\raisebox{2pt}{\tikz{\draw[-,red,dashed,line width = 1.5pt](0,0) -- (10mm,0);}}}
\newcommand{\blackline}{\raisebox{2pt}{\tikz{\draw[-,black,solid,line width = 1.5pt](0,0) -- (10mm,0);}}}

\begin{abstract}
The study of sound propagation in a uniform duct having a mean flow has many applications, such as in the design of gas turbines, heating, ventilation and air conditioning ducts, automotive intake and exhaust systems, and in the modeling of speech. In this paper, the convective effects of the mean flow on the plane wave acoustic field inside a uniform duct were studied using artificial neural networks. The governing differential equation and the associated boundary conditions form a constrained optimization problem. It is converted to an unconstrained optimization problem and solved by approximating the acoustic field variable to a neural network. The complex-valued acoustic pressure and particle velocity were predicted at different frequencies, and validated against the analytical solution and the finite element models. The effect of the mean flow is studied in terms of the acoustic impedance. A closed-form expression that describes the influence of various factors on the acoustic field is derived.
\end{abstract}

\section{Introduction} \label{Sec:1}
It is essential to study the effect of mean flow on the acoustic field in a uniform duct at the initial stages of product development in various industries, including aerospace, HVAC, and automobiles. Several researchers developed analytical formulations to study the effect of mean flow on the plane wave acoustics field inside the ducts of regular geometries\cite{Easwaran1992,Eisenberg1971,Karthik2000,Li2017,Surendran2017,Munjal1986}. For complicated geometries and higher-dimensional problems, there are several techniques, including finite element methods (FEM)\cite{Ihlenburg1995}, boundary element methods (BEM)\cite{Shen2007}, and computational fluid dynamics (CFD) tools\cite{Lourier2012}. With the advancement of computational power and the development of new algorithms, such as for automatic differentiation, machine learning techniques gained momentum in addressing the problems related to noise and vibration using exhaustive data collected from the experiments or simulations. These techniques have been used successfully in structural health monitoring, especially for bearing fault detection\cite{Sharma2018,Mol2023,Kane2019} and their classification\cite{Mufazzal2022}. 

Recently, physics-based neural networks gained momentum in solving differential equations governing complicated physical laws\cite{Raissi2019}. In the realm of neural networks, the problem of solving a differential equation subjected to a given initial and/or boundary conditions is posed as an optimization problem and is solved by treating the initial and/or boundary conditions as soft constraints\cite{Van2022}. Several research teams successfully demonstrated the capabilities of neural networks in predicting field variables in high-fidelity equations (e.g. Navier-Stokes)\cite{Amalinadhi2022,Ang2022} as well as low-fidelity equations\cite{Dung2023,Karali2021,Palar2023}. 

Attempts have been made to solve the wave equation and predict the acoustic state variable (acoustic pressure and particle velocity) in the time domain\cite{Alguacil2021,Borrel2021}. However, predicting the acoustic state variables in the frequency domain is not a trivial task. At higher frequencies, the network encounters a vanishing gradient problem\cite{Wang2021}. This is addressed by automatic weight update procedures\cite{Van2022,Wang2021,Basir2022,Maddu2022}. However, these procedures involve hyperparameter tuning which is to be carried out manually for each individual frequency considered for the analysis. In addition to this, the acoustic field becomes a complex-valued function in several cases, e.g., a case where both the acoustic field and mean flow coexist. This poses a problem during the optimization process, as the objective/loss function has to be always real-valued. These two main problems are addressed in the current work, and the results obtained from the neural network formulation are validated with the analytical solution at selected frequencies up to 2000 Hz. 

The paper is organized as follows: Section~\ref{Sec:2} describes the neural network formulation for predicting the acoustic field in the frequency domain with and without mean flow. In Section~\ref{Sec:3}, the results obtained from the neural network formulation are compared with the analytical solution, and the effect of mean flow on the acoustic field is discussed. The paper concludes with the final remarks in Section~\ref{Sec:4}.

\section{Neural Network Formulation} \label{Sec:2}
Let us consider the problem of finding a scalar field function $\psi(x)$ which satisfies the following differential equation
\begin{equation}
    \mathcal{D}[\,\psi\,] = 0, \qquad \forall\,\, x\in\Omega, \label{Eq:1}
\end{equation}
subjected to the constraints
\begin{equation}
    \psi(x) = g(x), \qquad \forall\,\,x\in \partial\Omega, \label{Eq:2}
\end{equation}
where $\mathcal{D}[\,\cdot \,]$ is the differential operator, $\Omega$ is the domain, $\partial\Omega$ is the boundary of the domain, and $g(x)$ is the source function associated with the boundary constraints. 

\begin{figure}[hbt!]
\centering
\includegraphics[scale=1]{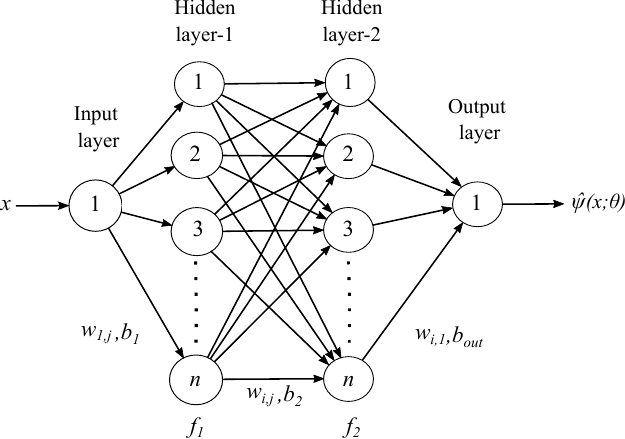}
\caption{\label{fig:FIG1}{Schematic diagram of a neural network architecture.}}
\end{figure}

According to the universal approximation theorem\cite{Hornik1989}, the function $\psi(x)$ can be approximated to any desired accuracy with a neural network $\hat{\psi}(x;\theta)$ provided that $\psi(x)$ is bounded and continuous in $\Omega$. Here, $\theta$ represents the parameters of the neural network, that is, weights ($w_{i,j}$) and biases ($b_i$), where $i=$ 1, 2, 3,..., $n_l-1$, $j=$ 1, 2, 3,..., $n$, $n_l$ is the number of layers, and $n$ is the number of neurons in each layer in the network. Figure~\ref{fig:FIG1} shows the schematic diagram of a neural network with two hidden layers and $n$ neurons in each layer. 

The function $\hat{\psi}(x;\theta)$ can be found by solving the following minimization problem\cite{Raissi2019}
\begin{equation}
\begin{aligned}
\min_{\theta} \quad & \mathcal{L}_f(x_f;\theta), \quad x_f\in\Omega \\
\textrm{s.t.} \quad & \mathcal{L}_b(x_b;\theta)=0, \quad x_b\in\partial\Omega \label{Eq:3}
\end{aligned}
\end{equation}
where
\begin{align}
    \mathcal{L}_f(x_f;\theta) &= \frac{1}{N_f}\sum_{i=1}^{N_f}\left\|\mathcal{D}[\,\hat{\psi}(x^{(i)}_f;\theta)\,]\right\|^2_2 \label{Eq:4}, \\
    \mathcal{L}_b(x_b;\theta) &= \frac{1}{N_b}\sum_{i=1}^{N_b}\left\|\hat{\psi}(x^{(i)}_b;\theta)-g(x^{(i)}_b)\right\|^2_2. \label{Eq:5}
\end{align}
Here, $N_f$ and $N_b$ are the number of randomly collocated points inside the domain and on the boundary, with $i$-th point represented by $x^{(i)}_f$ and $x^{(i)}_b$, respectively.

It is a constrained optimization problem. In general, it can be converted into an unconstrained problem using the Lagrange multiplier $\lambda$ as follows\cite{Raissi2019,Basir2022}
\begin{equation}
\min_{\theta} \quad  \mathcal{L}_f+\lambda\mathcal{L}_b.
\end{equation}
However, finding the appropriate $\lambda$ is not a trivial task, especially when the required output is an oscillatory function that depends on the frequency\cite{Wang2021}. Therefore, in this work, an alternative methodology known as the trial solution method is adopted, as it bypasses the need to use $\lambda$ in the optimization procedure\cite{Lagaris1998}. According to this method, a trial neural network solution $\hat{\psi}_t(x;\theta)$ is constructed in such a way that it \emph{exactly} satisfies the boundary constraints given in Eq.~(\ref{Eq:2}). The resulting unconstrained problem can be written as\cite{Lagaris1998}
\begin{equation}
\min_{\theta} \quad \mathcal{L}(x;\theta), \quad x=\{x_f,x_b\}, \label{Eq:6}
\end{equation}
where
\begin{equation}
    \mathcal{L}(x;\theta) = \frac{1}{N}\sum_{i=1}^{N}\left\|\mathcal{D}[\,\hat{\psi}_t(x^{(i)};\theta)\,]\right\|^2_2. \label{Eq:7}
\end{equation}
Here, $N$ is the number of randomly collocated points of the entire domain including the boundary, with each point represented by $x^{(i)}$. 
\subsection{Neural Network Formulation without Mean Flow}
When the mean flow is not present, the acoustic pressure inside a duct of length $L$ in the plane wave region can be obtained by solving the 1-D Helmholtz equation\cite{Morse1986}
\begin{equation}
    \left(\frac{d^2}{dx^2}+k^2\right)\psi(x)=0, \qquad x\in\left[0, L\right],
\end{equation}
subjected to the boundary conditions (constraints)
\begin{equation}
    \psi(0) = \psi_0; \qquad \psi(L) = \psi_L, \label{Eq:9}
\end{equation}
where $\psi(x)$ is the acoustic pressure, $k=2\pi f/c$ is the wavenumber, $f$ is the frequency and $c$ is the speed of sound.

The trial neural network $\hat{\psi}_t(x;\theta)$ which \emph{exactly} satisfies the boundary conditions mentioned in Eq.~(\ref{Eq:9}) can be constructed as follows\cite{Lagaris1998}
 \begin{equation}
    \hat{\psi}_t(x;\theta) = \phi_L\psi_0+\phi_0\psi_L+\phi_0\phi_L\hat{\psi}(x;\theta), \label{Eq:10}
\end{equation}
where
\begin{equation}
    \phi_L = \frac{L-x}{L}; \qquad \phi_0 = \frac{x}{L}. \label{Eq:11}
\end{equation}
Now, $\hat{\psi}_t(x;\theta)$ can be estimated by solving the following optimization problem
\begin{equation}
\min_{\theta} \quad \mathcal{L}(x;\theta), \quad x\in\left[0, L\right], \label{Eq:12}
\end{equation}
where
\begin{equation}
    \mathcal{L}(x;\theta) = \frac{1}{N}\sum_{i=1}^{N}\left\|\left(\frac{d^2}{dx^2}+k^2\right)\hat{\psi}_t(x^{(i)};\theta)\right\|^2_2. \label{Eq:13}
\end{equation}

By appropriately choosing the architecture of the network shown in Fig.~\ref{fig:FIG1}, the loss function $\mathcal{L}(x;\theta)$ can be minimized and the acoustic pressure in the absence of the mean flow can be predicted. Note that this formulation does not involve any hyperparameters which require manual tuning.



\subsection{Neural Network Formulation with Mean Flow} \label{Sec:3B}
When sound is propagating in a stationary medium, its speed ($c$) will be the same for the forward and backward wave components. However, when it is propagating inside a moving medium, the forward wave component will have an effective speed of $c+U$, while the backward wave component will have an effective speed of $c-U$, where $U$ is the mean flow velocity\cite{Easwaran1992}. This convective effect can be incorporated into the 1-D Helmholtz equation and the modified equation can be written as\cite{Li2017}
\begin{equation}
    \left[(1-M^2)\frac{d^2}{dx^2}-2jMk\frac{d}{dx}+k^2\right]\psi(x)=0, \qquad x\in\left[0, L\right], \label{Eq:14}
\end{equation}
where $M=U/c$, the mean flow Mach number and $j=\sqrt{-1}$.

The additional term appearing due to the presence of the mean flow makes the acoustic pressure $\psi(x)$, complex-valued. The neural network which is used for the case of stationary medium has the ability to predict the complex-valued acoustic pressure as well with minor modifications. 

Let the complex-valued acoustic pressure be
\begin{equation}
    \psi(x)=\psi^R(x)+j\psi^I(x), \label{Eq:15}
\end{equation}
where $R$ and $I$ in the superscript denote the real and imaginary parts, respectively. Substituting Eq.~(\ref{Eq:15}) in Eq.~(\ref{Eq:14}) yields
\begin{equation}
    \left[(1-M^2)\frac{d^2}{dx^2}-2jMk\frac{d}{dx}+k^2\right]\psi^R(x)+j\left[(1-M^2)\frac{d^2}{dx^2}-2jMk\frac{d}{dx}+k^2\right]\psi^I(x)=0 \label{Eq:16}
\end{equation}
Let us assume the same boundary conditions as in the case of stationary medium, i.e.,
\begin{align}
    \psi^R(0)+j\psi^I(0) &= \psi_0, \label{Eq:17} \\
    \psi^R(L)+j\psi^I(L) &= \psi_L. \label{Eq:18}
\end{align}
Equations (\ref{Eq:16}), (\ref{Eq:17}), and ((\ref{Eq:18})) can be written as two sets of governing equations, one for the real-part and the other for the imaginary-part, along with the associated boundary conditions as follows \\ \\
\underline{Equations associated with the real-part:}
\begin{align}
    (1-M^2)\frac{d^2\psi^R}{dx^2}+2Mk\frac{d\psi^I}{dx}+k^2\psi^R &= 0, \label{Eq:19}\\
                                  \psi^R(0) &= \psi_0; \qquad \psi^R(L) = \psi_L.
\end{align}
\underline{Equations associated with the imaginary-part:}
\begin{align}
    (1-M^2)\frac{d^2\psi^I}{dx^2}-2Mk\frac{d\psi^R}{dx}+k^2\psi^I &= 0, \label{Eq:21} \\
                                  \psi^I(0) &= 0; \qquad \psi^I(L) = 0.
\end{align}
The corresponding trial neural networks can be constructed as follows
\begin{align}
    \hat{\psi}_t^R &= \phi_L\psi^R(0)+\phi_0\psi^R(L)+\phi_0\phi_L\hat{\psi}^R, \\
    \hat{\psi}_t^I &= \phi_L\psi^I(0)+\phi_0\psi^I(L)+\phi_0\phi_L\hat{\psi}^I.
\end{align}
Now, the complex-valued acoustic pressure $\hat{\psi}_t=\hat{\psi}^R_t+j\hat{\psi}^I_t$ can be predicted by solving the following optimization problem
\begin{equation}
\min_{\theta} \quad \mathcal{L}^R(x;\theta)+\mathcal{L}^I(x;\theta), \quad x\in\left[0, L\right], \label{Eq:25}
\end{equation}
where $\mathcal{L}^R$ and $\mathcal{L}^I$ are the loss functions associated with the real and imaginary parts, respectively. They can be calculated as
\begin{align}
    \mathcal{L}^R &= \frac{1}{N}\sum_{i=1}^{N}\left\|(1-M^2)\frac{d^2\hat{\psi}^R_t(x^{(i)};\theta)}{dx^2}+2Mk\frac{d\hat{\psi}^I_t(x^{(i)};\theta)}{dx}+k^2\hat{\psi}^R_t(x^{(i)};\theta)\right\|^2_2, \label{Eq:26} \\
    \mathcal{L}^I &= \frac{1}{N}\sum_{i=1}^{N}\left\|(1-M^2)\frac{d^2\hat{\psi}^I_t(x^{(i)};\theta)}{dx^2}-2Mk\frac{d\hat{\psi}^R_t(x^{(i)};\theta)}{dx}+k^2\hat{\psi}^I_t(x^{(i)};\theta)\right\|^2_2. \label{Eq:27}
\end{align}

\begin{figure}[hbt!]
\centering
\includegraphics[scale=1]{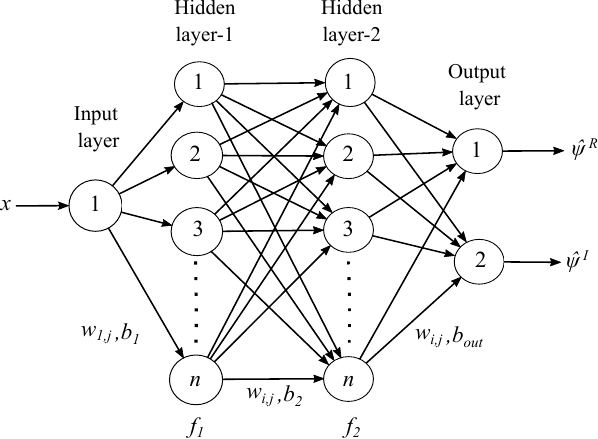}
\caption{\label{fig:FIG2}{Schematic diagram of the neural network architecture for complex-valued output.}}
\end{figure}

There is no need to create a separate neural network for each part of the equations. The neural network architecture which is used for the stationary medium can be used to predict the complex-valued acoustic pressure $\hat{\psi}=\hat{\psi}^R+j\hat{\psi}^I$ associated with the moving medium. This can be achieved by modifying the output layer
of the neural network such that it has two neurons, one to represent the real-part and the other to represent the
imaginary-part, as shown in Fig.~\ref{fig:FIG2}.

\section{Results and Discussion} \label{Sec:3}
\begin{figure}[h!]
\centering
\includegraphics[scale=1]{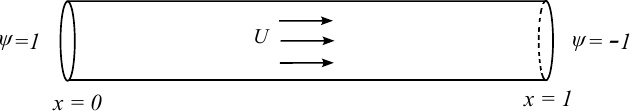}
\caption{\label{fig:FIG3}{Schematic diagram of the domain with boundary conditions.}}
\end{figure}

In the current work, a uniform duct of length $L=$ 1 m is considered for the study. The boundary conditions are assumed to be
\begin{equation}
    \psi(0) = 1; \qquad \psi(1)=-1.
\end{equation}
Figure~\ref{fig:FIG3} shows the schematic diagram of the domain with boundary conditions. The medium is assumed to be air and the speed of sound is taken as 340 m/s. The analysis is carried out till 2000 Hz, starting from 500 Hz, in steps of 500 Hz.

To predict acoustic pressure with and without mean flow, a feedforward neural network is constructed with 5 hidden layers ($n_l=7$) and 90 neurons in each hidden layer ($n=90$). The domain $\left[0, 1\right]$ is divided into $N=$ 10000 randomly collocated points. The biases of the network are initialized with zeros, and weights are initialized with He initialization (refer Appendix~A). The optimization of the loss function is performed using \emph{L-BFGS} algorithm with \emph{tanh} activation function. A total of 14000 iterations were performed with an optimal tolerance of 10$^{-5}$. These parameters are chosen according to those existing in the literature\cite{Raissi2019,Wang2021,Basir2022,Dung2023,Van2022}.

\subsection{Acoustic Pressure without Mean Flow}
\begin{figure}[h!]
\centering
\includegraphics[scale=1.3]{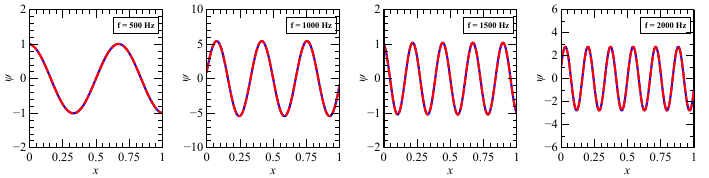}
\caption{\label{fig:FIG4}{Acoustic pressure without mean flow at different frequencies: \protect\blueline true solution, \protect\redline predicted solution.}}
\end{figure}

Figure~\ref{fig:FIG4} shows the comparison of the acoustic pressure predicted by the neural network formulation (predicted solution) against the analytical solution (true solution ) at different frequencies. Refer to Appendix B for the analytical solution. It can be observed that the predicted solution is in good agreement with the true solution. In other words, the neural network is able to learn the underlying physics behind the governing differential equation at the collocated points of the domain. 

The relative error between the predicted solution and the true solution is calculated as per the below formula and is used as a correlation metric throughout this work. 
\begin{equation}
    \delta \psi = \frac{\sqrt{\displaystyle\sum_{i=1}^{N_t} |\hat{\psi}_t(x^{(i)};\theta)-\psi(x^{(i)})|^2}}{\sqrt{\displaystyle\sum_{i=1}^{N_t}|\psi(x^{(i)})|^2}}. 
\end{equation}
Here, $N_t$ is the total number of linearly spaced test points in the domain $\left[0, 1\right]$, which are used to predict the solution using neural network formulation as well as to calculate the analytical solution. In the current study, $N_t$ is taken as 500. The relative error obtained at different frequencies is tabulated in Table~\ref{tab:table1}. From the results, it is evident that the neural network formulation serves as an effective tool to predict the acoustic field governed by physical laws.
\begin{table}[ht]
\centering
\caption{\label{tab:table1} Relative error at different frequencies without mean flow}
\begin{tabular}{cc}
\toprule \toprule
\textbf{Frequency (Hz)} & \textbf{$\delta\psi$} \\
\midrule \midrule
500 &  8.5538$\times$ 10$^{-07}$\\ \midrule
1000 &  3.6359$\times$ 10$^{-05}$\\ \midrule
1500 &  7.7487$\times$ 10$^{-06}$\\ \midrule
2000 &  5.1628$\times$ 10$^{-05}$\\ 
\bottomrule \bottomrule 
\end{tabular} \\ [10pt]
\end{table}

\subsection{Acoustic Pressure with Mean Flow}
\begin{figure}[h!]
\centering
\includegraphics[scale=1.2]{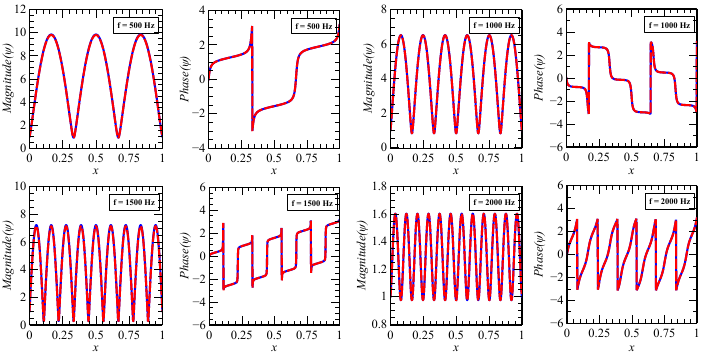}
\caption{\label{fig:FIG5}{Acoustic pressure with mean flow at different frequencies: \protect\blueline true solution, \protect\redline predicted solution.}}
\end{figure}

As discussed in Section~\ref{Sec:3B}, the additional convective term due to the presence of mean flow makes the acoustic pressure complex valued. Figure~\ref{fig:FIG5} shows the comparison of the acoustic pressure in terms of the magnitude and phase at different frequencies for a uniform mean flow of $U=$ 34 m/s, i.e., $M=$ 0.1. Refer to Appendix C for the analytical solution. It can be observed that the neural network formulation is capable of predicting both magnitude and phase with reasonable accuracy. Table~\ref{tab:table2} shows the relative error in magnitude and phase at different frequencies. It is to be noted that the neural network architecture which is used for the case of \emph{without mean flow} has been used here. 

\begin{table}[ht]
\caption{\label{tab:table2} Relative error in magnitude and phase at different frequencies with mean flow}
\centering
\begin{tabular}{cccc}
\toprule \toprule
\textbf{Frequency (Hz)} & \textbf{$\delta\psi_{mag}$} & \textbf{$\delta\psi_{phase}$} \\ 
\midrule \midrule
500 &  0.0044 & 0.0005\\ \midrule
1000 &  0.0014 & 0.0001\\ \midrule
1500 &  0.011 & 0.0016\\ \midrule
2000 &  0.0008 & 0.0002\\
\bottomrule \bottomrule 
\end{tabular} \\ [10pt]
\end{table}
In the neural network formulation, the additional term which introduces convective effects splits the governing equation (Eq.~(\ref{Eq:14})) into two coupled equations (Eqs.~(\ref{Eq:19}) and (\ref{Eq:21})). This introduces mathematical complexity into the training process. The optimization has to be performed in such a way that both loss functions (Eqs.~(\ref{Eq:26}) and (\ref{Eq:27})) are minimized simultaneously. In other words, the network has to identify the shared parameters that satisfy the equations associated with both real and imaginary parts, simultaneously. The results reveal that the chosen network architecture and training parameters are able to handle such complexity efficiently (with a maximum error of 1.1\% in the magnitude).

\subsection{Effect of Mean Flow on the Acoustic Field}
As discussed earlier, the mean flow affects the forward and backward propagating waves differently. This effect has been studied in terms of the acoustic impedance for different flow conditions. The particle velocity $\xi(x)$ is estimated from the acoustic pressure $\psi(x)$ using the momentum equation\cite{Munjal2014}. 

In the absence of mean flow, the particle velocity can be estimated from the acoustic pressure as follows
\begin{equation}
    \xi = -\frac{1}{j\omega\rho}\frac{d\psi}{dx},
\end{equation}
where $\omega$ is the angular frequency, and $\rho$ is the density of the medium (air). As $\psi$ is approximated with a neural network $\hat{\psi}$, it can be differentiated with respect to $x$, and the approximated $\hat{\xi}$ can be estimated from the above relation. However, when the mean flow is present, the momentum equation gives the following relationship between the acoustic pressure and particle velocity
\begin{equation}
    jk\xi+M\frac{d\xi}{dx}=-\frac{1}{\rho c}\frac{d\psi}{dx},
\end{equation}
where $\rho$ is the density of air. 

Now, the estimation of particle velocity from the acoustic pressure is not a trivial task, as there is an additional gradient term in terms of $\xi$ in the relation. This is where the neural network formulation offers an advantage over the traditional methods. Since $\xi(x)$ is also a continuous and bounded function, it can be approximated with another neural network $\hat{\xi}(x;\Tilde{\theta})$ and can be estimated by minimizing the loss function
\begin{equation}
    \mathcal{L}_\xi = \mathcal{L}_\xi^R + \mathcal{L}_\xi^I,
\end{equation}
where 
\begin{align}
    \mathcal{L}_\xi^R &= \frac{1}{N}\sum_{i=1}^{N}\left\|M\frac{d\hat{\xi}^R_t(x;\Tilde{\theta})}{dx}-k\hat{\xi}^I_t(x;\Tilde{\theta})+\frac{1}{\rho c}\frac{d\hat{\psi}^R_t(x;\Tilde{\theta})}{dx}\right\|^2_2,  \\
    \mathcal{L}_\xi^I &= \frac{1}{N}\sum_{i=1}^{N}\left\|M\frac{d\hat{\xi}^I_t(x;\Tilde{\theta})}{dx}+k\hat{\xi}^R_t(x;\Tilde{\theta})+\frac{1}{\rho c}\frac{d\hat{\psi}^I_t(x;\Tilde{\theta})}{dx}\right\|^2_2.
\end{align}
Here, $\hat{\xi}^R_t$ and $\hat{\xi}^I_t$ are the real and imaginary parts of the particle velocity $\hat{\xi}_t$. Note that the parameters of $\hat{\xi}_t$ are different from those of $\hat{\psi}_t$. Since the particle velocity is calculated from the predicted acoustic pressure using the trial neural network $\hat{\psi}_t$, the same subscript notation $(\,\cdot \,)_t$ is also carried forward to it.
\begin{figure}[h!]
\centering
\includegraphics[scale=1.3]{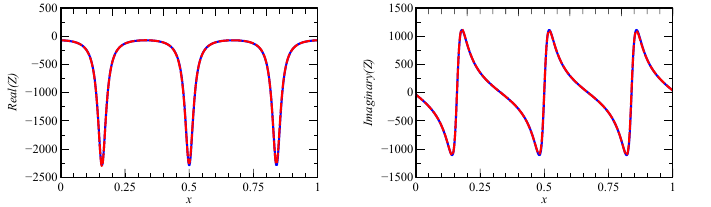}
\caption{\label{fig:FIG6}{Real and imaginary parts of acoustic impedance for $f=$ 500 Hz and $M=$ 0.1: \protect\blueline true solution, \protect\redline predicted solution.}}
\end{figure}

From the predicted acoustic pressure and particle velocity, the acoustic impedance ($Z=\psi/\xi$) is calculated and plotted in Fig.~\ref{fig:FIG6} in comparison with the analytical solution (Appendix C) for the frequency of 500 Hz. The density of air $\rho$ is considered to be 1.225 kg/m$^3$ in the calculations. It can be observed that the predicted solution is able to capture the positions of peaks and valleys at par with the true solution. The relative errors are observed to be 1.5\% and 1.9\% for the real and imaginary parts, respectively.

\begin{figure}[h!]
\centering
\includegraphics[scale=0.85]{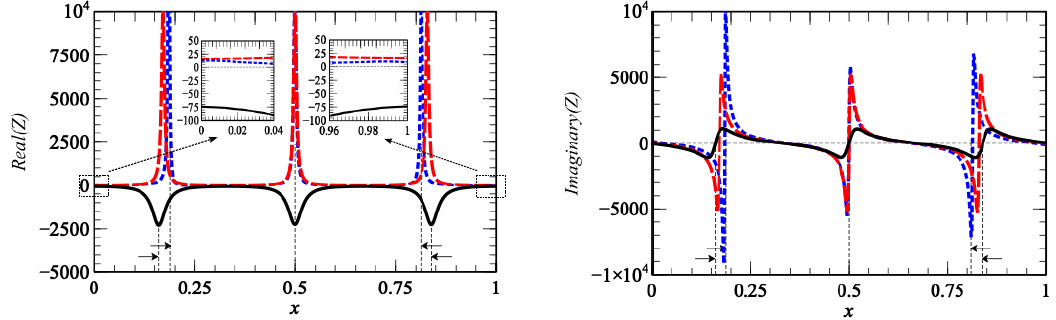}
\caption{\label{fig:FIG7}{Effect of mean flow on the acoustic impedance: \protect\blackline $M=$ 0.1, \protect\redline $M=$ 0.2, \protect\bluedotline $M=$ 0.3.}}
\end{figure}

Figure~\ref{fig:FIG7} shows the comparison of the real and imaginary parts of the impedance for different mean flow conditions: $M=$ 0.1, 0.2, and 0.3. At the chosen frequency (500 Hz), the system has three velocity nodes. It can be observed that the mean flow has a significant effect on the position of these nodes. The increase in mean flow shifts the left-node toward the right side and the right-node toward the left side, keeping the middle node intact. Thus, the symmetry of the problem is preserved. 

\begin{figure}[h!]
\centering
\includegraphics[width=0.6\textwidth]{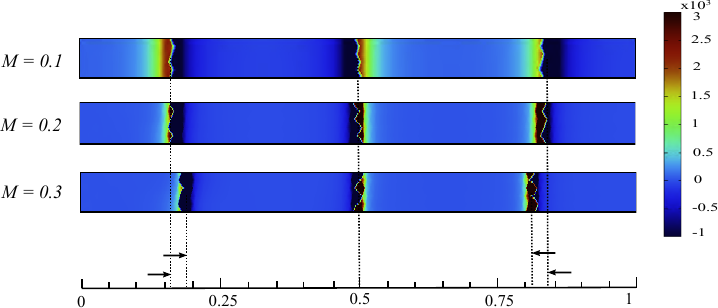}
\caption{\label{fig:FIG8}{Real-part of the acoustic impedance obtained from 2-D FEM for different flow conditions.}}
\end{figure}
This phenomena can be validated with the two-dimensional finite element method (2-D FEM). Figure~\ref{fig:FIG8} shows the real-part of the acoustic impedance obtained from the 2-D FEM model developed using COMSOL Multiphysics commercial solver\cite{Comsol2022}. A rectangular domain of length 1 m and width 40 mm is created. The width of the domain is chosen in such a way that only plane waves exist inside the domain within the chosen maximum interested frequency $f_{max}$. In other words, $f_{max}$ is chosen such that
\begin{equation}
    f_{max} < f_c = \frac{c}{2W},
\end{equation}
where $f_c$ is the cut-off frequency, and $W$ is the width of the domain. The domain is discretized into a number of triangular elements. The maximum size of the element is kept below one-sixth of the minimum wavelength ($c/f_{max}$). The domain is assigned with the same medium properties, i.e., $c=$ 340 m/s and $\rho=$ 1.225 kg/m$^3$. The following boundary conditions
\begin{align}
    \psi(0,y) &= 1, \quad y \in [0, W] \\
    \psi(1,y) &= -1, \quad y \in [0, W]
\end{align}
are applied on the left and right boundaries, respectively. The top and bottom boundaries are assumed to be acoustically rigid, i.e.,
\begin{align}
    \xi(x,0) &= 0, \quad x \in [0, L]\\
    \xi(x,W) &= 0, \quad x \in [0, L].
\end{align}
The model is solved for different flow conditions using linearized Euler equations in the frequency domain. From the acoustic pressure $\psi$ and particle velocity $\xi$, the acoustic impedance $Z=\psi/\xi$ is calculated, and its real-part is compared for $M=$ 0.1, 0.2, and 0.3, in Fig.~\ref{fig:FIG8}

The mean flow not only alters the position of the velocity nodes but significantly influences the behavior of the real-part of the impedance. It can be observed from Fig.~\ref{fig:FIG7} that real-part of the impedance for $M=$ 0.1 is negative throughout the domain, whereas it is completely positive for $M=$ 0.2 and 0.3. It indicates that the mean flow introduces out-of-phase behaviour in the real-part of the impedance $Re(Z)$ at $f=$ 500 Hz.  The reason for this behaviour can be analyzed by considering the real-part of the impedance at any point in $\left[0, 1\right]$ as its behaviour (positive/negative) is the same throughout the domain. For simplicity, consider the impedance value at the left boundary of the domain, that is, $x=$ 0. From the plane wave theory, it can be written as
\begin{equation}
    Z(0) = \frac{\psi(0)}{\xi(0)} = \rho c \frac{\psi_0}{A_c-B_c},  
\end{equation}
where $A_c$ and $B_c$ are the constants and their values are evaluated in Appendix C.

Upon substituting the values of $A_c$ and $B_c$ in the above impedance expression and performing a few algebraic operations, the expression for the real-part of the impedance at $x=$ 0 can be written as
\begin{equation}
    Re(Z(0)) = \frac{2\rho c\psi_0\psi_L\left(\cos k_c^+-\cos k_c^-\right)}{\left[2\psi_L-\psi_0\left(\cos k_c^++\cos k_c^-\right)\right]^2+\left[\psi_0\left(\sin k_c^+-\sin k_c^-\right)\right]^2}.
\end{equation}
It can be inferred from the above expression that the \emph{sign} of $Re(Z(0))$ depends on the values of the numerator and denominator. The denominator of $Re(Z(0))$ is always positive and finite at $x=$ 0. Furthermore, the value of $2\rho c$ in the numerator is always positive. Therefore, the \emph{sign} of $Re(Z(0))$ depends on the \emph{sign} of
\begin{equation}
    s:=\psi_0\psi_L\left(\cos k_c^+-\cos k_c^-\right),
\end{equation}
where $k_c^+ = k/(1+M)$ and $k_c^- = k/(1-M)$ are the convective wavenumbers of the forward and backward wave components, respectively. The analysis reveals that the \emph{sign} of $Re(Z(0))$ is not just a function of the mean flow ($M$), it also depends on the boundary conditions ($\psi_0,\psi_L$) and the wavenumber ($k$). The \emph{sign} of $Re(Z)$ depends on the \emph{sign} of $s$ as follows
\begin{equation}
    Re(Z)  \begin{cases}
                >0 &\text{when} \quad s>0,\\
                <0 &\text{when} \quad s<0.
                \end{cases}
\end{equation}
For the current problem under consideration $\psi_0\psi_L=$ -1. Hence, the \emph{sign} change for $Re(Z)$ takes place when
\begin{equation}
    \cos k_c^+-\cos k_c^- = 0.
\end{equation}
The non-trivial solution of the above equation is as follows
\begin{equation}
    k_c^+ = 2\pi m\pm k_c^-,
\end{equation}
where $m=$ 1, 2, 3, and so on. It can be easily verified that for the frequency of 500 Hz and $m=$ 1, the \emph{sign} change for $Re(Z)$ takes place at $M=$ 0.14 in the mean flow Mach number range $\left[0.1, 0.3\right]$, that is, $Re(Z)<$ 0 for $M\in$ $\left[0.1, 0.14\right]$, and is positive for the rest of the range. This underlying physics is effectively captured by the neural network formulation. Note that for any other value of $m$, the \emph{sign} change takes place at $M>$ 0.3 which is outside the region of interest.

 
\section{Conclusion}\label{Sec:4}
The acoustic field in a uniform duct in the presence of mean flow is predicted using artificial neural networks in the plane wave region. Two cases, without and with mean flow, were studied. The acoustic field in both cases is predicted with reasonable accuracy. A concatenated neural network architecture containing two distinct networks, the first one is to predict the acoustic pressure, and the second one is to predict the particle velocity from the Euler momentum equation is proposed. The acoustic impedance computed from the predicted results is compared with the analytical solution, and good agreement is observed. It is observed that the mean flow shifts the position of velocity nodes and introduces out-of-phase behaviour in the real-part of the acoustic impedance. The analysis reveals that, apart from the mean flow, frequency, and boundary conditions are also responsible for the out-of-phase behaviour. A closed-form analytical expression comprising all the influential factors is derived. 

The main drawback of physics-informed neural networks, especially when used to predict the acoustic field in the frequency domain, is the requirement of tuning the hyperparameters and/or loss function weights. These have to be tuned for every frequency either manually or by automatic means. In addition, predicting complex-valued field variables is not a trivial task, as the loss function has to always be a real-valued function to perform the optimization process. These difficulties can be bypassed by adapting the formulation presented in the current manuscript. The results presented throughout the manuscript are obtained using a single neural network architecture. In addition, the method presented in the current manuscript is not limited to one-dimensional cases. It can be extended to higher dimensions to perform the analysis beyond the plane wave region. This shows the ability of the neural networks to serve as a tool besides traditional numerical techniques such as FEM, BEM, CFD tools, etc. in the near future.

\bibliographystyle{ieeetr}
\bibliography{references}

\appendix
\section{He Initialization} \label{Appendix:A}
The weights $\mathbf{w}_i$ in a layer $i=$ 2, 3, 4, ..., $n_l-1$ are initialized with He initialization using the formula\cite{He2015}
\begin{equation}
    \mathbf{w}_i = \sqrt{\frac{2}{S_{i-1}}}\times randn(S_i,S_{i-1}),
\end{equation}
where \emph{randn} is a function to generate normally distributed random variables of size $S_i\times S_{i-1}$ with mean $\mu=$ 0 and variance $\sigma=$ 1. In other words, $S_i\times S_{i-1}$ random variables are to be generated with the density function:
\begin{equation}
    p(z) = \frac{1}{2\sqrt{\pi}}e^{-z^2/2},
\end{equation}
where $z$ is a random variable.

\section{Analytical Solution without Mean Flow} \label{Appendix:B}
In the absence of mean flow, the acoustic pressure in the plane wave region can be obtained by solving the following 1-D Helmholtz equation\cite{Morse1986}
\begin{equation}
    \left(\frac{d^2}{dx^2}+k^2\right)\psi(x)=0, \qquad x\in [0,\,L], \label{Eq:B1}
\end{equation}
subjected to the boundary conditions
\begin{equation}
    \psi(0)=\psi_0; \qquad \psi(L)=\psi_L.
\end{equation}
The governing equation (Eq.~(\ref{Eq:B1})) is an ordinary differential equation with constant coefficients. It can be solved by assuming a solution of the form
\begin{equation}
    \psi = Ce^{\lambda x}, \label{Eq:B3}
\end{equation}
where $\lambda$ is an eigenvalue. Upon substituting Eq.~(\ref{Eq:B3}) in Eq.~(\ref{Eq:B1}) yields the solution 
\begin{equation}
    \psi(x)=A\cos(kx)+B\sin(kx), \label{Eq:B4}
\end{equation}
where $A$ and $B$ are constants that can be calculated from the boundary conditions. Substituting these boundary conditions in Eq.~(\ref{Eq:B4}), the analytical solution can be written as 
\begin{equation}
    \psi(x)=\psi_0\cos(kx)+\left[\frac{\psi_L-\psi_0\cos(kL)}{\sin(kL)}\right]\sin(kx). \label{Eq:B5}
\end{equation}

\section{Analytical Solution with Mean Flow} \label{Appendix:C}
The 1-D acoustic pressure with convective mean flow effect can be obtained by solving the following equation\cite{Li2017}
\begin{equation}
     \left[(1-M^2)\frac{d^2}{dx^2}-2jMk\frac{d}{dx}+k^2\right]\psi(x)=0, \qquad x\in\left[0, L\right], \label{Eq:C1}
\end{equation}
subjected to the boundary conditions
\begin{equation}
    \psi(0)=\psi_0; \qquad \psi(L)=\psi_L
\end{equation}
Since Eq.~(\ref{Eq:C1}) is also an ordinary differential equation with constant coefficients, by following the similar procedure as in Appendix B, the general solution can be obtained as
\begin{equation}
    \psi(x)=A_ce^{-jk_c^+x}+B_ce^{jk_c^-x}, \label{Eq:C3}
\end{equation}
where $k_c^+=k/(1+M)$ and $k_c^-=k/(1-M)$. The constants $A_c$ and $B_c$ can be obtained from the boundary conditions as follows
\begin{align}
    A_c &= \frac{\psi_L-\psi_0e^{jk_c^-}}{e^{-jk_c^+}-e^{jk_c^-}}, \\
    B_c &= \psi_0-A_c.
\end{align}
Similarly, the particle velocity $\xi$ can be assumed to take the form as
\begin{equation}
    \xi(x)=C_ce^{-jk_c^+x}+D_ce^{jk_c^-x}, 
\end{equation}
and the constants $C_c$ and $D_c$ can be calculated from the momentum equation\cite{Munjal2014} 
as 
\begin{align}
    C_c &= \frac{A_c}{\rho c}, \\
    D_c &= -\frac{B_c}{\rho c}.
\end{align}

\section*{Acknowledgments}
The authors acknowledge the support received from the Department of Science and Technology, and Science and Engineering Research Board (SERB), Government of India towards this research.

\end{document}